\documentclass[article,twocolumn,showpacs,superscriptaddress,amsmath,amssymb]{revtex4}
\usepackage{cancel}
\usepackage{maybemath}
\usepackage{xcolor}
\usepackage{graphicx}

\begin{document}

\title{The Tachyon Nexus:\\  
An Educational Resource on Tachyons and Time Travel
}
\author{Robert Ehrlich}
\affiliation{George Mason University, Fairfax, VA 22030}
\email{rehrlich@gmu.edu}
\date{\today}

\begin{abstract}
A web site ``The Tachyon Nexus'' (no relation to Tachyon Nexus Inc) is described.  This web site (http://ehrlich.physics.gmu.edu) includes much reliable information about the controversial subjects of time travel and faster-than-light particles known as tachyons mostly from a physics point of view.  This compendium of various resources should be of great value both to students seeking to learn about these subjects or to high school teachers or college professors wishing to include them in their teaching on special or general relativity.
\end{abstract}
\maketitle
There have been numerous articles\citep{1,2,3,4,5,6,7,8,9}on the hypothetical tachyon in physics education journals including one in The Physics Teacher.\citep{4} In fact, the very first article on the subject in 1962 appeared in an education journal,\citep{9} since it was regarded as too speculative for the research journal to which it was originally submitted.  Most theorists today regard faster-than-light tachyons as being non-existent for many reasons, not the least of which is the possibility they might allow you to send messages back in time, and they might even make the universe unstable!  Despite these obnoxious properties there is another type of tachyon that is commonly used in field theories, including the one behind the Higgs boson.  Of course, the existence of the faster-than-light tachyon is a matter for experiment to decide, despite the opinions of theorists.  There have been previous searches for tachyons, and one claim in 2011 for faster-than light neutrinos was later retracted when they discovered a loose cable in the timing circuit.  However, such negative searches cannot rule out tachyons any more than negative searches for ET can rule out the existence of intelligent aliens, so the status of tachyons continues to be unsettled as of now.  

Even if tachyons are in the hypothetical realm, suspended between fact and fiction, there is considerable value in bringing science fiction topics into physics courses.\citep{10,11,12,13,14} This is especially true in courses for liberal arts majors, and for that sort of science fiction topics that might become science fact someday.  By adding elements of controversy and speculation to teaching about our well-established theories like relativity, we show students that physics is a living enterprise, and much remains unsettled.  Hypothetical entities like tachyons or wormhole time machines can be more stimulating for students to read about than well-established entities like the Higgs boson because they can imagine themselves making the great discovery that establishes their existence, and who knows maybe they will.  Tachyons are an especially tantalizing possibility because Einstein ruled out faster-than light particles in his first relativity paper, and who wouldn't like to disprove something Albert said.  

There are further reasons both for bringing science fiction into physics courses, and also for physics students and faculty to read science fiction on their own. It is noteworthy, for example, that both the author and Gerald Feinberg – the very physicist who gave the name tachyons to faster-than-light particles were aware of the topic way through science fiction stories.\citep{15}   Morever astrophysicist Richard Gott has shown that science fiction writers have been surprisingly prescient in exploring ideas that have proven to be topics of serious investigation of time travel among scientists.
\citep{15a}  As a prime example, ten years before relativity H. G. Wells wrote his epic novel The Time Machine, in which the protagonist discusses what would later be called worldlines through four-dimensional spacetime.  There is no evidence that Hermann Minkowski got these very ideas from reading or hearing about Well's novel, but it is certainly within the realm of the possible.

As is well known, the topics of time travel and faster-than-light speeds are closely connected.  This connection is well-established in popular imagination, through A. H. Reginald Butler's famous limerick,\citep{16} several popular television series, and many movies.  One-way time travel to the future is certainly possible based on the time dilation effect, even though it will be a while before spaceships achieve the speeds needed to have a space traveling twin age very much less than their stay-at-home twin.  Time travel to the past, however, like tachyons remains controversial in physics, but there have been many papers written about it, and some of the world's leading physicists, including Stephen Hawking have remained open to its possibility, despite his original skepticism.\citep{17}  

Many students will try to learn about subjects like tachyons and time travel by browsing the web, which can be quite risky given the many questionable web sites dealing with such highly controversial subjects.  Web browsing to learn about tachyons is particularly risky, given the considerable amount of nonsense connected to so-called "tachyon energy" and "tachyon healing," and the many commercial products that have given the very word a bad name.  Aside from the reasonably accurate Wikipedia page on tachyons, there is nearly a complete absence of web sites that treat the subject in a serious way.  Similarly a search for serious web sites dealing with time travel in physics that go beyond an article on the subject yields very little, again beyond a wikipedia page.  

For this reason, I have created "The Tachyon Nexus," a (not always serious) web site where both physics students and physics teachers can learn more about the two subjects of tachyons and time travel.  The web site includes slide presentations at several levels, many frequently asked questions (with answers) about tachyons and time travel, the up-to-date status of current research on tachyons, and links to various other web sites.  It also includes a blog where users can launch discussions about issues related to tachyons amongst themselves.  As someone who has done research on the subject of tachyons, I believe that I have built a site that is authoritatively correct, accessible to both students and teachers, and not excessively biased.  I do, however, plead guilty in believing that the world of physics may be in for one very big surprise in the coming year or two concerning the existence of tachyons.

\end{document}